\documentclass[12pt,a4paper]{article}

\usepackage{apalike}
\usepackage{graphicx}
\usepackage{geometry}

\begin{document}

\title{Towards Self-organizing Bureaucracies}
\author{Carlos Gershenson \\
Centrum Leo Apostel, Vrije Universiteit Brussel\\
Krijgskundestraat 33 B-1160 Brussel, Belgium\\
cgershen@vub.ac.be \ http://homepages.vub.ac.be/\symbol{126}cgershen}
\maketitle

\begin{abstract}
This paper proposes self-organization as a method to improve the efficiency and adaptability of bureaucracies and similar social systems. Bureaucracies are described as networks of agents, where the main design principle is to reduce local ``friction" to increase local and global ``satisfaction". Following this principle, solutions are proposed for improving communication within bureaucracies, sensing public satisfaction, dynamic modification of hierarchies, and contextualization of procedures. Each of these reduces friction between agents (internal or external), increasing the efficiency of bureaucracies. ``Random agent networks" (RANs), novel computational models, are introduced to illustrate the benefits of self-organizing bureaucracies.
\end{abstract}

\section{Introduction}

Bureaucracies can be found in governments, corporations, and other social
institutions. They have social goals and responsibilities that are achieved
by a division of labor that is usually hierarchical. Examples of
bureaucracies can be seen with tax collection systems, immigration services,
and steering of educational and academic institutions. The efficiency of a
bureaucracy is related to the fulfillment of its goals. Thus, it would be
desirable to increase functional efficiency in bureaucracies. Ideally, such
a system could be designed to reach maximal efficiency. In practice, as most
people have experienced, this is far from being the case \cite{Weber1968}.
Corruption, rigidity, and delays are just few examples of obstacles that
hamper efficiency in bureaucracies. It would be na\"{\i}ve to aim for
perfect bureaucracies, but certainly the efficiency of actual ones can be
improved.

One approach would consist of \emph{optimizing} the bureaucratic
functionality, e.g. \cite{HofackerVerschera2001}. This approach can provide
good solutions if the function or goal of the organization does not change
considerably, i.e. when a problem domain is static. However, the world is
changing at accelerating rates. Changes cause the shifting of the optimum of
a system. And in some cases, the behaviour of the institution itself changes
the optimum \cite{Kauffman2000}. Thus, a wiser approach would be to design bureaucracies that
are able to \emph{adapt} \cite{Holland1975} to changing situations. Instead
of attempting to predict all the functionality beforehand, an organization
could adapt to the changing demands of its environment.

Cybernetics and systems theory proposed some of the first solutions in this
direction already a few decades ago, e.g. \cite{Beer1966}. The Cybersyn
project in Chile was even partially implemented, but was cut short by the
1973 military coup \cite{MillerMedina2005}. However, this approach is still
not widely used in practice, probably because it requires alternative ways
of thinking. It is always easier to solve problems for static domains than
for dynamic environments.

Organization science has developed several concepts that are useful for
improving the self-organization and adaptation of bureaucracies. Noting the
cognitive limits of decision makers \cite{March1978,Simon1982,CyertMarch1992}
tells us that individuals will not be able to make perfect decisions. On one
hand, the cognition necessary to solve complex tasks can be distributed \cite%
{Hutchins1995,WeickRoberts1993}. On the other hand, organizations need to be
able to adapt to unpredictable events \cite{Carley1997}. Organizational
learning is one type of adaptation that has been widely studied \cite%
{LevittMarch1988}. Also, computational organization theory \cite%
{CarleyPrietula1994} and agent based modelling, e.g. \cite%
{EpsteinAxtell1996,Axelrod2005}, have aided in the understanding of the
complexity inherent to organizations \cite%
{AndersonEtAl1999,Anderson1999,LevinthalWarglien1999,Lissack1999,AxelrodCohen2000}

Following this line of research, this work suggests methods that improve the efficiency of bureaucracies via self-organization. In the next section, a notion of self-organization is introduced, for then presenting general ideas for the design of self-organizing systems. In the following sections, different aspects of bureaucracies, and self-organizing improvements are proposed, namely in the areas of communication, sensors, hierarchies, and context. Afterwards, random agent networks (RANs) are defined to model bureaucracies. Computational experiments with RANs illustrate the benefits of self-organization for improving the performance of abstract bureaucracies.

\section{Self-organization}

Self-organization \cite{Heylighen2003sos} has been used as a principle in
many domains such as computer science and robotics \cite{Kohonen2000}, the
Internet \cite{BollenHeylighen1996}, and traffic light control \cite%
{Gershenson2005}, just to name a few. In organization science,
self-organization has been studied as a phenomenon, e.g. \cite[p.233]%
{Comfort1994,Morgan1996}. The goal of this work is to use it conceptually as
a \emph{tool} to improve the efficiency of bureaucracies.

Even when it has been widely studied and applied in many domains,
self-organization is a concept difficult to define \cite%
{GershensonHeylighen2003a}. Nevertheless, for practical purposes, a notion
can suffice: A system \emph{described} as self-organizing is one in which
elements \emph{interact} in order to achieve \emph{dynamically} a global
function or behavior \cite{Gershenson2006}. A classical example can be seen
with flocks of birds, or schools of fish: there is no leader in the group,
and all individuals follow local rules, interacting with their neighbors,
and only this produces coherent global behaviour. In a similar way, local
rules of interaction can be designed, to produce dynamically robust and
adaptive behaviours. In this work, some of such rules are proposed, as a way
to improve the performance of bureaucracies.

\section{Designing Self-organizing Systems}

Organizations can be modelled as systems of information processing agents 
\cite{Radner1993,VanZandt1998,DeCanioWatkins1998}. An agent is a description
of an entity that \emph{acts} on its environment. They could also be
described as \emph{cognitive} systems \cite{Gershenson2004}. Thus, not only
people in a bureaucracy can be described as agents, but also departments,
ministries, and whole bureaucracies. Agents can have goals \cite{Simon1964},
that are described by an observer. Agents can be said to be cognitive
because they need to ``know" which actions to take to reach
their goals. The ``satisfaction" of the agents will be
related to the achievement of their goals. Thus, a description of a
bureaucracy can be made in terms of agents trying to fulfill goals to
increase their satisfaction. The public can also be described as an agent or
several agents, interacting (externally) with the bureaucracy.

However, the satisfaction of one agent (e.g. a clerk) can be in conflict
with the satisfaction of another agent (e.g. the minister). It can be argued
that decreasing the ``friction" or interference of agents at
one level (e.g. personal level), i.e. how one agent decreases the satisfaction of another agent,
will result in an increase of satisfaction at the higher level (e.g.
ministry level) \cite{Gershenson2006,HelbingVicsek1999}. If one agent at the lower level is not
satisfied, i.e. does not accomplish its goals, the satisfaction of the
system may be reduced. However, if all agents at the lower level fulfill
their goals, then the satisfaction of the system should be maximal. This is
a (useful) tautology because the goals of agents are described by observers
according to the desired function of the system. Notice that this is
different from implying that increasing the satisfaction of agents at one
level will always lead to an increase of satisfaction at the higher level.
The key difference lies in the mediation of conflictive goals to increase
satisfaction. Similarly, we can speak about ``negative
friction", or synergy \cite{Haken1981}, where the behaviour of one agent
increases the satisfaction of another agent. Certainly, not only friction
should be minimized, but also synergy maximized. Different ways in which
this can be achieved are discussed in \cite{Gershenson2006}. Friction
reduction and synergy promotion will be always useful, since all actors,
internal or external, will be benefitted. For example, the easier it is to
pay taxes, the more people that will pay them. This benefits the state (more
money collected) and taxpayers (less time lost). Certainly, as in most
social systems, a problem will remain when it comes to measuring
satisfaction. This will be discussed in Section \ref{secSensors}.

The goals of a firm can be easily related to its profits \cite{VanZandt1998}%
. However, the goals of a bureaucracy are related to its particular
function. This function can be codetermined by the state, by the public,
and by the bureaucracy itself. Thus, there is yet no general way to measure
the performance of a bureaucracy. Efficiency could be a way of evaluating a
bureaucracy, but there is the same measurement problem with efficiency: it
will differ accordingly to the particular bureaucracy. Still, a lack of
explicit descriptions of function, efficiency, or satisfaction are not a
limit for speaking about the goals of a bureaucracy. It should just be
considered that these can change depending on the behaviour of the
bureaucracy itself.

In order to adapt to unpredictable changes, bureaucracies require a certain
flexibility. Changes should be made, but the function needs to be preserved.
Robustness is required \cite{Jen2005}, so that adaptive changes do not
prevent the bureaucracy from reaching its goals. The main idea to guide
changes is the following: First, detect how each agent affects satisfaction
of others. Then, implement changes to minimize friction and promote synergy.
This can be achieved by reinforcement: behaviours that have proven
themselves inefficient should be avoided, and beneficial ones should be
promoted.

Before implementing radical changes, computer simulations should be used 
\cite{Axelrod2005}. These will be useful for detecting possible flaws in the
changes planned, or simply to improve them. Moreover, the changes themselves
can be explored with the aid of computer simulations, since it is not
obvious in every case what should be done, as the complexity of
bureaucracies exceeds our predictive capabilities.

The changes could be introduced gradually and with a certain redundancy to
compare the benefits and disadvantages of the new methods with the previous
ones.

In the following sections particular aspects of bureaucracies are explored,
suggesting possible improvements within each domain.

\section{The Role of Communication}

Communication between agents can be classified in two categories:
synchronous and asynchronous \cite{DesanctisMonge1999}.

\emph{Synchronous} communication occurs when the agents involved in the
process are responding at the same time. There is immediate feedback between
speaker and listener, so that a dialogue can be established continuously.
Examples of this are verbal communication, telephone conversations, video
conferencing, and IRC (Internet relay chat). The advantage of this mode of
communication is that dialogues can be resolved without interruption. The
disadvantage is that all participants need to coordinate to participate in
the process.

\emph{Asynchronous} communication occurs when the agents involved do not
participate simultaneously in the process. There is a delayed feedback
between agents, so that dialogues are interrupted depending on the length of
the transmission delay. Examples include post, telegraph, telex, fax,
e-mail, and instant messaging. The tradeoffs of this mode of communication
complement those of synchronous communication: on the positive side
asynchronous communication allows exchanges without coordination required,
but on the negative side the communication can be delayed. Technological
development has reduced transmission delay, enabling asynchronous
communication to depend only on the constraints of the agents.

In a bureaucracy, different agents need to communicate to satisfy the goals
of the system. Thus, communication delays can be seen as a type of friction
between agents. The faster the communication takes place, the better it will
be for the system. Thus, synchronous communication might be preferred to
enable quick responses. However, this would imply a great coordination
effort, since agents usually perform other activities apart from
communicating. It could be quite possible that an agent would be busy with
other matters to have a synchronous exchange. Then it seems that
asynchronous communication would be preferred, since one agent can send a
message and keep on working on other matters while the response arrives.
Then the question would be: how can asynchronous communication be facilitated?

As mentioned above, one great improvement is given by technology. Being able
to send documents electronically instead of postally reduces the delay of
message transmission from the scale of days to the scale of seconds.
Certainly, organizations have exploited this opportunity, and worries about
security have been solved with digital signatures. Still, it is a common
practice in several bureaucracies to handle paper documents, even when they
must be sent across continents, as it is the case e.g. with the Mexican
foreign services.

\begin{figure}[htbp]
\begin{center}
\includegraphics[height=10cm] {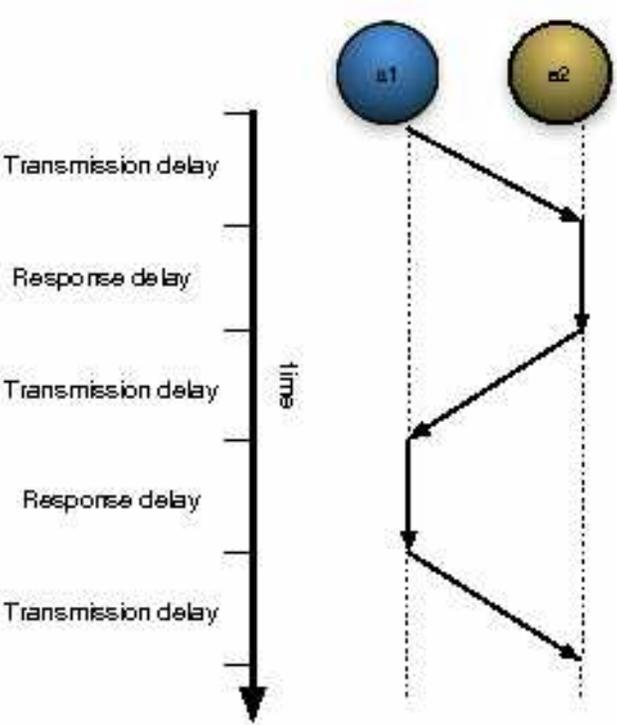}
\caption{Asynchronous communication. Technology has reduced transmission
delays, and can help to detect and decrease response delays.}
\label{asynchComm-delays}
\end{center}
\end{figure}

However, the adoption of electronic means of communication can do much more
than reducing the transmission delay of messages. Analyzing the times when a
message is sent and when it is replied can provide very useful information,
namely that of response delay (see Figure \ref{asynchComm-delays}). This can be used to detect
bottlenecks: If one agent (individual or department) takes too long in
replying requests, the work of other agents might be delayed as well, as in
a production chain. Resources could then be reassigned in real time to
overcome the bottleneck, giving priority to the agent with the response
delay. In fact, we can say that a delay in response causes friction to other
agents, since they need the feedback to reach their goals. Thus, a
bureaucracy could self-organize by modifying in real time its own structure,
once it is known where friction is coming from. Solutions can vary depending
on the precise nature of the delay: assign more individuals to a department,
replace individual(s), or reorganize departments. Like this, efficiency of
the bureaucracy would be improved. It would be self-organizing, because the
changes are dictated by the behaviour of the bureaucracy itself. The changes
would imply \emph{learning} in the organization from its experience, while
enabling it to adapt constantly to changes of its environment.

\begin{figure}[htbp]
\begin{center}
\includegraphics[height= 10cm]
{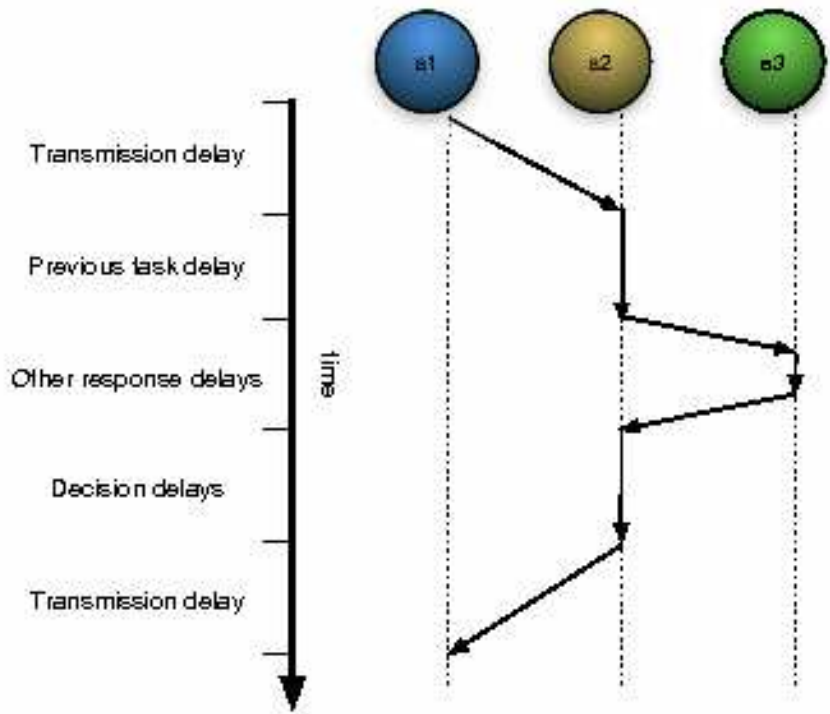}
\caption{Response delay can be decomposed in previous task delay, other
response delay, and decision delay.}
\label{asynchComm-responsedelay}
\end{center}
\end{figure}

The response delay would depend on several factors: decision delay (the time
it takes the agent to respond), delay from previous tasks (the time it takes
an agent to start making a decision), and delay from other responses (the
time it takes other agents to respond to the agent's requirements) (see Figure \ref{asynchComm-responsedelay}).
Each of these delays should be taken into account while modifying the
bureaucracy.

Another benefit of electronic logs is that they can be used to provide
accountability of decisions, as it was the case for the Enron e-mail dataset 
\footnote{See http://www.cs.cmu.edu/\symbol{126}enron/}. The workload of individual agents or
departments could also be measured by the number of requests they send or
receive, considering the decision delays required by each request. Finally,
a visualization of the interactions within the bureaucracy (who communicates
with who) could provide insights to improve its design, for example
detecting redundant agents or interactions, or creating ``shortcuts" between
agents that communicate frequently via other agents.

\subsection{Decision Delays}

Technology has also aided in the reduction of decision delays. Electronic databases provide instant information, while in a physical repository a clerk has to search an archive for documents. The role of the clerk is taken by software
in an electronic database. Similarly, monotonous decisions can be taken by
computer systems near instantly, reducing decision delays. An example of
this can be seen with bank credit evaluation \cite{HandHenley1997}, where a
computer system can give instant decisions on wether to give credit to an
applicant or not. Similar methods could be used to make instant decisions to
judge e.g. visa applicants or prospective students. Turning decisions into
computer systems will reduce decision delays, thus reducing friction,
improving communication, and increasing the bureaucracy efficiency. This is
precisely one of the directions agent technology is taking \cite%
{AgentRoadmap2005}, using notions of negotiation, trust, and reputation to
facilitate the coordination of electronic decision makers. Also research and
technology applied to e-government \cite{LayneLee2001} and computer aided
decision making \cite{TurbanAronson1997,Stahl2006} will improve the
performance of bureaucracies.

Such a hybrid scenario, where humans and agents interact in an organization,
has been described with the term ``cognitive stigmergy" \cite{RicciEtAl2006}.
Stigmergy is used to describe systems, such as insect colonies, that exploit
their environment to communicate and coordinate \cite{TheraulazBonabeau1999}%
. In a similar way, computer systems can be used as an environment to
facilitate the communication, coordination, and decision of agents.

\section{The Role of Sensors}

\label{secSensors}

In the previous section, reduction of friction within the bureaucracy was
discussed. In this section, reduction of friction between the bureaucracy
and its environment, namely the public, will be discussed.

Much research has been made in decision making, e.g. \cite{Simon1976}. It is clear that
without proper sensors there will not be enough information to make proper
decisions. Still, even with simple sensors, a system can obtain much
information from its surroundings. An example can be seen with people who
perceive their environment with a walking stick, sensing by pressure only
one point in space. Integrating information in time, they are able to obtain
relevant information to navigate through complex areas. Nevertheless,
complex sensors can reduce the complexity of a decision making process, by
``digesting" relevant information. Therefore, bureaucracies
should aim at developing fine sensors to perceive and digest information
relevant to their goals. In any case, without proper sensors, proper
decisions cannot be made.

One element that facilitates the sensing process is public participation,
since people themselves digest and feed the information to the bureaucracy.
However, many people are reluctant to participate in such processes, since
they do not see any benefit for it, while it takes some of their time. An
alternative would be to reconstruct public opinion from a limited set of the
population, as polls have been doing, and novel methods could improve, e.g. 
\cite{RodriguezSteinbock2004}. Here, only improvement of sensors that do not
require public participation will be discussed.

In order to measure the efficiency of a bureaucracy, sensors should be used.
A popular variable related to this efficiency is public satisfaction: if
the public is happy with the services provided by the bureaucracy, then its
efficiency can be assumed. Polls have been used to measure public
satisfaction, but demand a certain effort from public and resources to
design and analyze them. Also, they cannot measure all possible mishaps.

Thus, bureaucracies should develop sensors for public satisfaction
that do not require public participation. This could be done measuring the
public attention delay, which would be the sum of the waiting delay (how
much time a person needs to queue) and the procedure delay (how much time a
person needs to interact with the bureaucracy). Another indicator would be
the frequency of interaction, namely how many times the same person needs to
interact with the bureaucracy. These delays can be considered as friction
between the bureaucracy and the public, and should be minimized. Both the
public and the bureaucracy will be satisfied if they need to interact with
each other as few as possible (low interaction frequency), and each of these
interactions takes as little time as possible (low public attention delay).
Like this, the precise places where bottlenecks arise, and for which cases,
can be detected, and measures can be taken. For example, if a procedure for a special
type of license takes consistently more time than others, this procedure should
be revised and adapted.

One could argue that bureaucracies do not need to care for the public, since
they can be considered as monopolies. But the tendency towards
improving bureaucratic services refutes this argument. For example,
e-government practically eliminates the waiting delays. It is beneficial for
political parties in office to improve bureaucratic services to increase
public satisfaction, and thus more votes for the next election. Natural
selection will give better chances of survival to political parties that
satisfy pubic demands. Certainly, this can only happen in countries with a
certain degree of political diversity. Otherwise, indeed the state would be
a monopoly.

\section{The Role of Hierarchies}

Hierarchies are certainly useful for organizations \cite{HelbingEtAl2006}. A problem might arise when these are too
rigid and changes are necessary for adaptation. Moreover, when several
aspects should be dealt by a bureaucracy, it might be that one agent should
be above another in some aspect (e.g. logistics), whereas in a different
aspect the opposite might be the case (e.g. legal advice).

Ashby's law of requisite variety \cite{Ashby1956} tells us that a system
needs to have enough variety to respond to the variety of its environment
(the word variety here could be substituted for the word complexity). A hierarchy
could also be necessary for coping with environmental complexity \cite{Aulin1979}. Multiscale analysis \cite{Bar-Yam2005} is a formal tool that can
be used to determine when a hierarchy is required. Basically, if the
complexity of an environment cannot be coped by individual agents, these need to aggregate and coordinate to cope collectively. The organizational
relations between agents lead naturally to hierarchies.

\begin{figure}[htbp]
\begin{center}
\includegraphics[width= 10cm]
{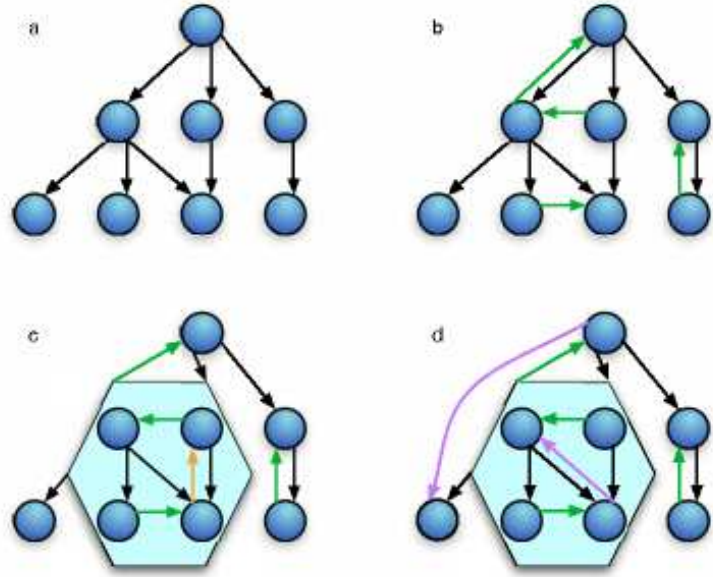}
\caption{Hierarchy represented as a network. a) Strict
hierarchical network. b) As interactions and dependencies increase, strict
hierarchy is broken, but still far from homogeneous. c) Modules can be
created when too many interactions cause delays. d) ``Shortcuts" can be made to avoid intermediaries. Links can also be removed or modified,
as the network adapts.}
\label{NetEvol}
\end{center}
\end{figure}

To visualize hierarchies, bureaucracies can be represented as
networks \cite{Strogatz2001,Newman2003}, where each node represents an agent
(at a particular scale), and edges represent interactions between agents.
Certainly, a network representing a bureaucracy will not be homogenous, since
different roles are taken by different agents. A hierarchical bureaucracy
can also be represented as a network (Figure \ref{NetEvol}a). As the complexity (variety)
demanded by the bureaucracy's environment increases, the diversity of roles
and interactions also augments. A solution would be to increase the number
of agents, but this would lead to longer communication delays. A better
alternative would be to increase the interaction types between the existing
agents, to avoid the introduction of new actors while coping with the
required complexity. These new interactions might change the strictly
hierarchical nature of the bureaucracy. However, even when the bureaucracy
might be highly distributed, a certain hierarchy will always be found,
simply because the network is not random nor homogeneous, i.e. there will
always be agents with more weight in the network's function than others
(Figure \ref{NetEvol}b).

In a system where too many agents need to interact at once, such as the
European Union with its current twenty five members, the complexity of the
interactions may be too difficult to manage. \emph{Modularity} can help in
coping with the complexity \cite{Simon1996}. Following the EU example, it
will be less complicated if some decisions are made e.g. by five groups of
five countries, and then these five groups discuss a final decision, than
having all members discussing at once. This is because the decision delays
of each agent add up, since agents (in theory) need to listen to other
agents before making a decision. In the modular case, discussions can go in
parallel, so it would take five decision delays for the first round, and
five for the second round, ten in total. In the plain case, it would take
twenty five decision delays to have a discussion. Certainly, too many
modules would also create delays. A balance should be sought where agents
can make decisions and interact as efficiently as possible. What is
important is to note that modularity in a network also implies a certain
hierarchy (Figure \ref{NetEvol}c). The size of modules will also be limited by the cognitive abilities of the agents \cite{Miller1956}, so in principle each human agent should not have more than seven interactions. Common sense would tell us that other types of agent should also keep the number of interactions low.

A desirable property of bureaucratic networks will be that they have a
``small world" topology \cite{WattsStrogatz2001}. This means that most interactions between agents will not need
many intermediaries. This is important for information transfer, again, to
reduce communication delays. Simply ensuring that messages do not need to
pass through several agents before reaching their destination will result in
a small world effect, because like this the agents that need to be connected
will be connected. If the same message needs to pass from agent A to agent
B, to finally reach agent C, it might be worthwhile to simplify and do a
shortcut from A to C (Figure \ref{NetEvol}d). This same idea was posed by \cite{BollenHeylighen1996} to improve website navigation by dynamically creating
direct links between pages that users reached via other pages.

In this scenario, a bureaucratic hierarchy is dynamic and changing when
necessary, adapting to changes of its environment directed by friction reduction. Again, these changes can
be said to be self-organizing, because the restructuration comes from within
the institution, directed by its own dynamics.

\section{The Role of Context}

In order to cope with complexity, any organization will try to simplify
procedures. Abstracting from several instances, details can be omitted, and
a uniform approach can be taken to deal with new instances, internal and
external. For example, the public tends to be treated uniformly. This makes
sense in cases when there are not many differences between individuals, e.g.
to apply for a passport: every citizen has a name, address, etc. However,
when a uniform approach is used for cases where there is diversity in the
public, difficulties may arise. A single template cannot predict beforehand
all cases, and usually makes simple cases complicated. An example can be
seen with certain tax declarations, that include very specific sections that
only few people must fill in, but are delivered to everybody, even if they
just need to sign and declare that they had no income. And changes in the
taxation policies can make it complicated enough to pay somebody to fill in
a declaration. Rather than including all possible cases in a single form, a
more reasonable approach would be to \emph{contextualize} the situation, providing
individual solutions for specific cases. Electronic media make this
feasible, by generating instant options depending on the current
circumstances.

Contextualizing interactions will reduce frictions, internal and external,
because both agents involved in the interaction would be benefited if delays
are reduced by removing considerations that do not apply for the current
situation. Certainly, too much contextualization can be counterproductive,
since agents need to learn how to deal with each case. If every case
requires new decisions, then expertise will not be able to improve the
performance of agents.

What could be done is to categorize contexts into common occurring
categories, by using one of many well known techniques for automatic
classification or clustering. Returning to the tax declaration example,
people who filled in similar parts of the form can be automatically
classified into a contextual category, such as pensioners or unemployed. Like
this, a system can find automatically which contexts are common, and what measures should
be taken only for those contexts. Since this would be a continuous process,
new contextual categories can arise and old ones may disappear. The
advantage is that these changes are lead by the usage of the
bureaucracy itself, satisfying its demands.

\section{A Toy Model: Random Agent Networks}

To have a better feeling of the usefulness of the ideas described so far, a simple computational model can be used to measure the performance of abstract bureaucracies, represented as ``random agent networks" (RANs). This model, partly inspired by random Boolean networks \cite{Kauffman1969,Kauffman1993,Wuensche1998,AldanaEtAl2003,Gershenson2004c}, tries to make as few assumptions as possible about the structures of bureaucracies.

A RAN consists of $N$ nodes. A node represents an agent, which could represent a person, a department, or a ministry. Each agent $i$ solves a task. To do so, it sends requests to $K_{i}$ other agents, which can be called ``dependencies" or connections. The dependencies of every agent are chosen randomly at the beginning of a simulation, not to assume any structure. Once the agent receives a response from all its dependencies, the task is complete. However, the dependencies might receive several requests from several agents. Thus, they store requests in a queue, which they attend in a first-come, first-served basis\footnote{For simplicity, in the model dependencies do not propagate: requests from queues are answered in one time step. In real bureaucracies, some of these requests might require further requests to further dependencies.}. Time is also abstracted, so agents take one time step to send requests (transmission delay), one time step to answer one request from the queue (decision delay), and one step to integrate the responses and complete a task (decision delay). Once a task is complete, agents start a new task. Agents respond requests from their queue only when they are expecting responses from their own dependencies.

The performance of the network can be measured by the number of tasks it is able to complete. Thus, the time ``wasted" by agents while waiting for responses from their dependencies (response delay) and having an empty queue should be minimized.

Now, there are many possible ways of randomly assigning dependencies to agents. The simplest would be \textbf{homogeneous}, where each agent has exactly $K$ dependencies chosen randomly. Following a \textbf{normal} probability distribution, every agent will have on average $K$ dependencies, so some agents will have more and some will have less. A more natural distribution would be \textbf{scale-free} \cite{Aldana2003}, where few agents have several dependencies, and most agents have not so many\footnote{More precisely, the number of dependencies for each agent is generated with the probability distribution $P(x)=(\gamma - 1)x^{-\gamma}$}. Intuitively, a special topology where every agent receives the same number of tasks should obtain the best performance, so that workloads are distributed equitably, not allowing request queues to grow for particular agents. A (non random) topology where every agent connects to K neighbours, similar to cellular automata, fulfills this requirements. This topology can be called \textbf{symmetric}.

The RAN model was implemented in a computer simulation written in the Java programming language. The reader can try the simulation and download the source code via the website http://rans.sourceforge.net

\begin{figure}[htbp] 
   \centering
   \includegraphics[width=6in]{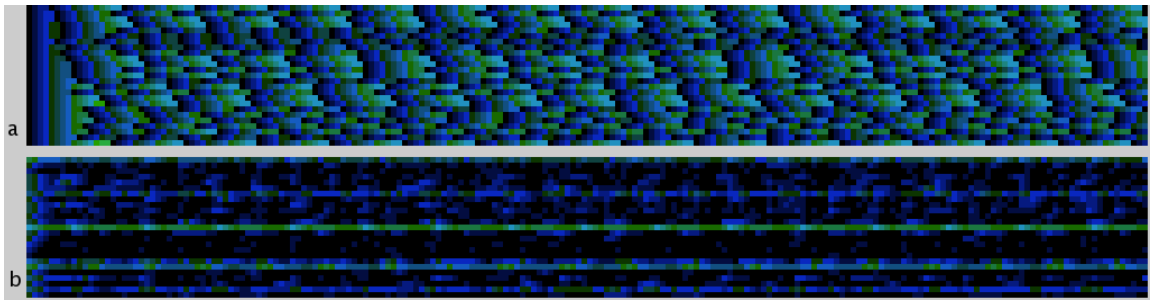} 
   \caption{Dynamics for a random agent network of $N=25$, $K=5$ with homogeneous topology for 200 time steps. a) Response delays. b)Queue lengths. Lighter colours indicate higher values. The initial state is the leftmost columns, and the subsequent columns show the temporal evolution of the RAN.}
   \label{RANdyn}
\end{figure}

As an initial state, all agents send requests to their dependencies. Afterwards, agents are updated sequentially each time step, i.e. there are $N$ updates per time step. The behaviour of the network converges to a periodic or quasi-periodic pattern, i.e. an attractor. Interesting parameters to observe are the response delays (how long it takes an agent to complete a task, which is determined by how quickly its dependencies are able to process its requests) and queue lengths (how many requests an agent has yet to process). An example of the dynamics of these parameters can be seen in Figure \ref{RANdyn}.

Except for the symmetric topology, for any values of $N$ and $K$, the task queues seem to converge typically to a power law distribution: there are few long ones (bottlenecks), and many short ones.

\subsection{Using self-organization to improve performance}

If we see the satisfaction of agents in terms of the tasks they are able to complete, the satisfaction will be lower when the response delay is higher. Agents with longer queues cause more friction than others, because they cause a high response delay on the agents that are dependent on them, i.e. they have a high previous task delay. Thus, a natural way of reducing friction would be by restructuring the network in such a way to reduce the longest queues.

\begin{figure}[htbp] 
   \centering
   \includegraphics[width=3in]{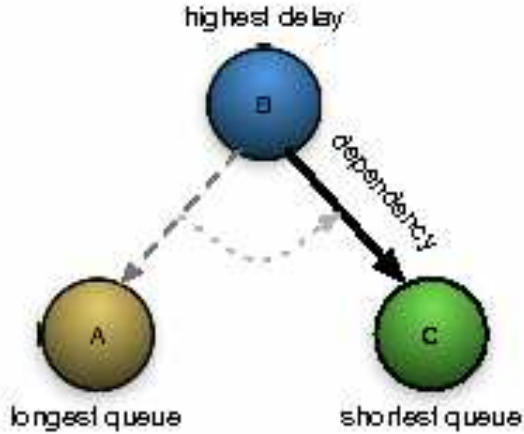} 
   \caption{Self-organization mechanism: The agent with highest delay ($B$) restructures its dependency from the agent with longest queue ($A$) to the one with shortest ($C$).}
   \label{SelfOrgMech}
\end{figure}

A very simple criterium achieves this. To restructure a RAN, the agent with the maximum queue average ($A$) is detected. Then, the agent with a maximum response delay ($B$) that has as a dependent the agent with the longest queue changes its dependency to the agent with the shortest queue ($C$). This mechanism is illustrated in Figure \ref{SelfOrgMech}. In many cases, the agent with highest response delay ($A$) has the agent with longest queue ($B$) as its dependency. This is natural, since $B$ will be able to complete its task only when $A$ reaches the request after processing its queue. It is obvious that changing the dependency of $B$ from $A$ to $C$ will reduce the response delay. What is not obvious is the precise effect that this will have on the network performance.

\subsection{Simulation Results}

To compare the scale-free with the other topologies, this was normalized to have a total number of dependencies in the network very close to $N*K$.\footnote{More precisely, $\gamma=2.48$ was used, since networks with this value have similar properties to $K=2$ \cite{Aldana2003}. Then, the probability was multiplied by $K/2$, to normalize.} The networks with normal topology also were checked to have a comparable number of dependencies, since networks with less dependencies are able to process more tasks.

\begin{figure}[htbp] 
   \centering
   \includegraphics[width=6in]{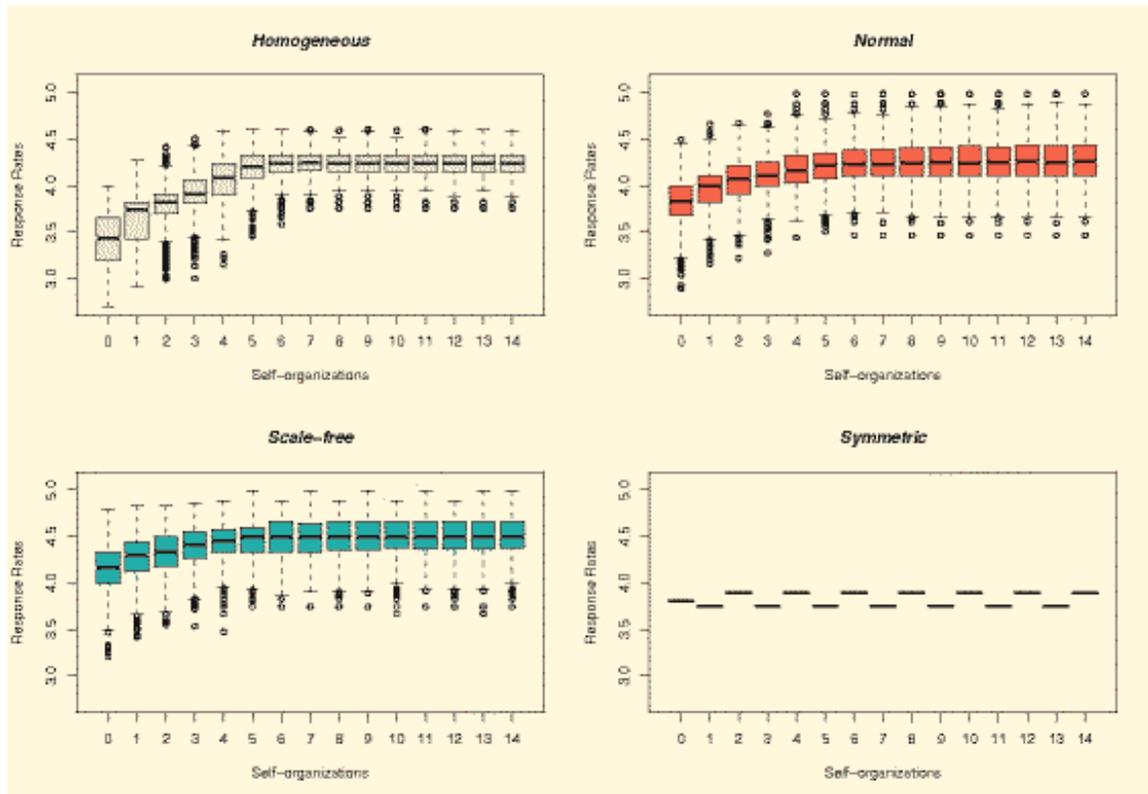} 
   \caption{Results for $N=15$, $K=1$.}
   \label{N15K1}
\end{figure}

Simulation runs were performed for different parameter values. For a network size of $N=15$, for each topology, 1000 RANs were created, and their response rate (average tasks completed per time step) was plotted as the self-organization was iteratively applied each 1000 time steps. For $K=1$ (Figure \ref{N15K1}), the scale free topology performs even better than the symmetric one. This is because there is usually only one or few nodes with lots of dependencies, and many nodes with few or no dependencies. The latter ones are able to complete their tasks quickly, since they have little or no interference from the delays of other nodes, and this enables them to respond quickly to the demands of the former ones. Note that there is already a certain hierarchy and modularity inherent in this configuration. As self-organization restructures the RANs, the non-symmetric topologies increase their performance, and after few reorganizations, the homogeneous and normal topologies also perform better than the symmetric. These two have initially bad performance because randomly some nodes are dependencies of more than one node, while others are dependency of none. This creates queues that affect all the nodes that share busy dependencies. By changing the dependency to idle agents, the idle agents are still able to complete their own tasks, while serving as dependencies of other nodes, and reducing the overall friction in the network.

\begin{figure}[htbp] 
   \centering
   \includegraphics[width=6in]{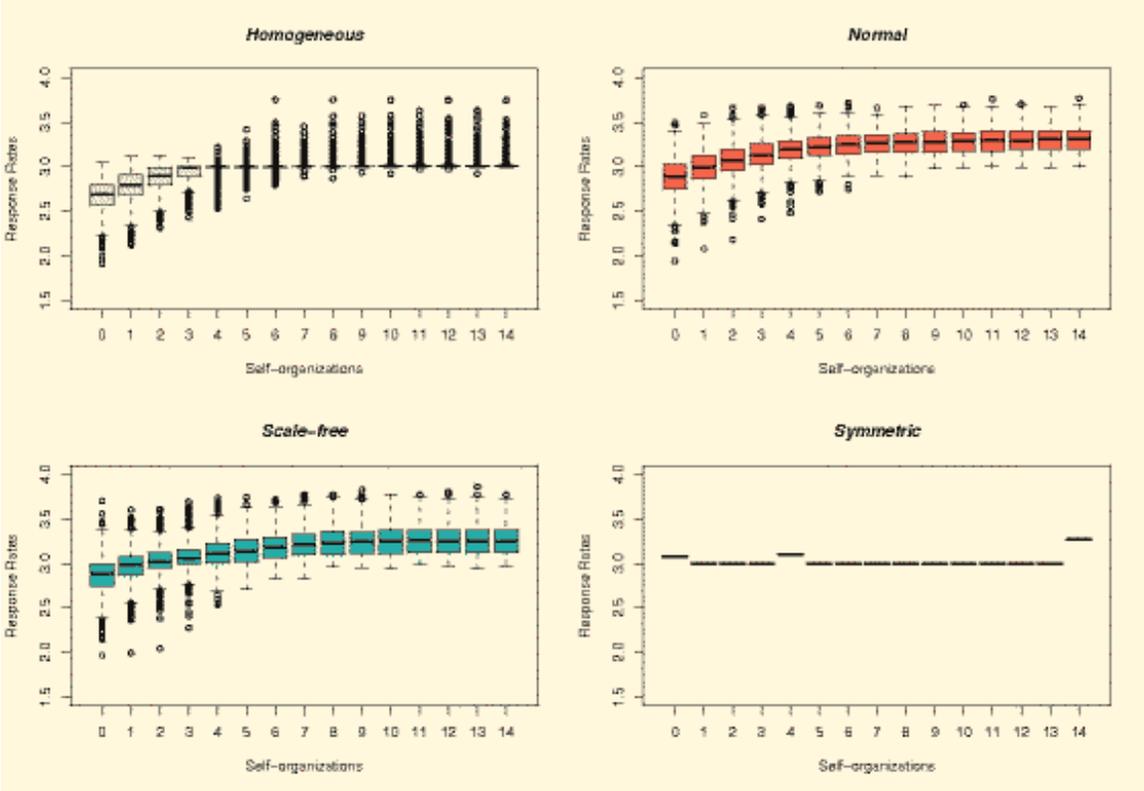} 
   \caption{Results for $N=15$, $K=2$.}
   \label{N15K2}
\end{figure}

\begin{figure}[htbp] 
   \centering
   \includegraphics[width=6in]{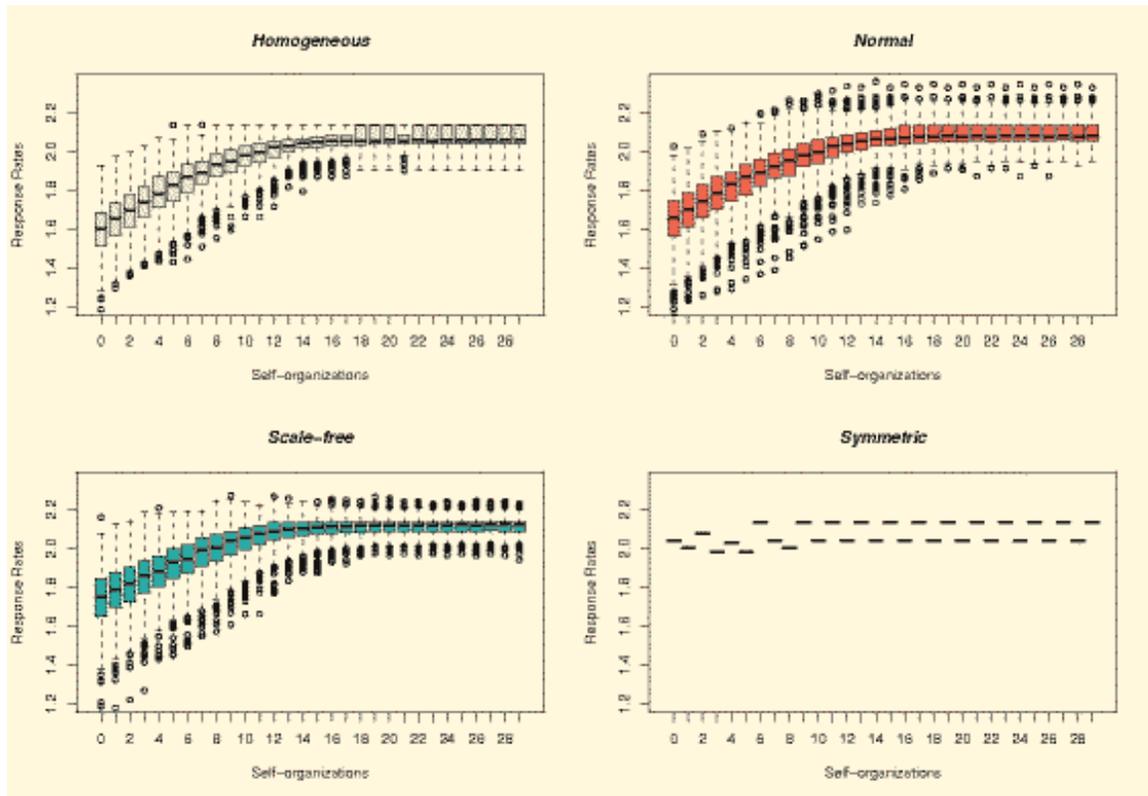} 
   \caption{Results for $N=15$, $K=5$.}
   \label{N15K5}
\end{figure}

For $K=2$ (Figure \ref{N15K2}), before self-organization, the symmetric topology performs the best. After few self-organizations, the homogeneous topology achieves the same performance, while the normal and scale free surpass it. This is because of the same reason explained above: if some nodes have several dependencies while these have few, overall they are able to process more tasks than if every node has the same number of dependencies: the nodes with few dependencies are able to process their own tasks quickly, and to respond to the agents with many dependencies promptly, creating certain hierarchy and modularity at the same time. A similar case is seen for the case when $K=5$ (Figure \ref{N15K5}).

\begin{figure}[htbp] 
   \centering
   \includegraphics[width=6in]{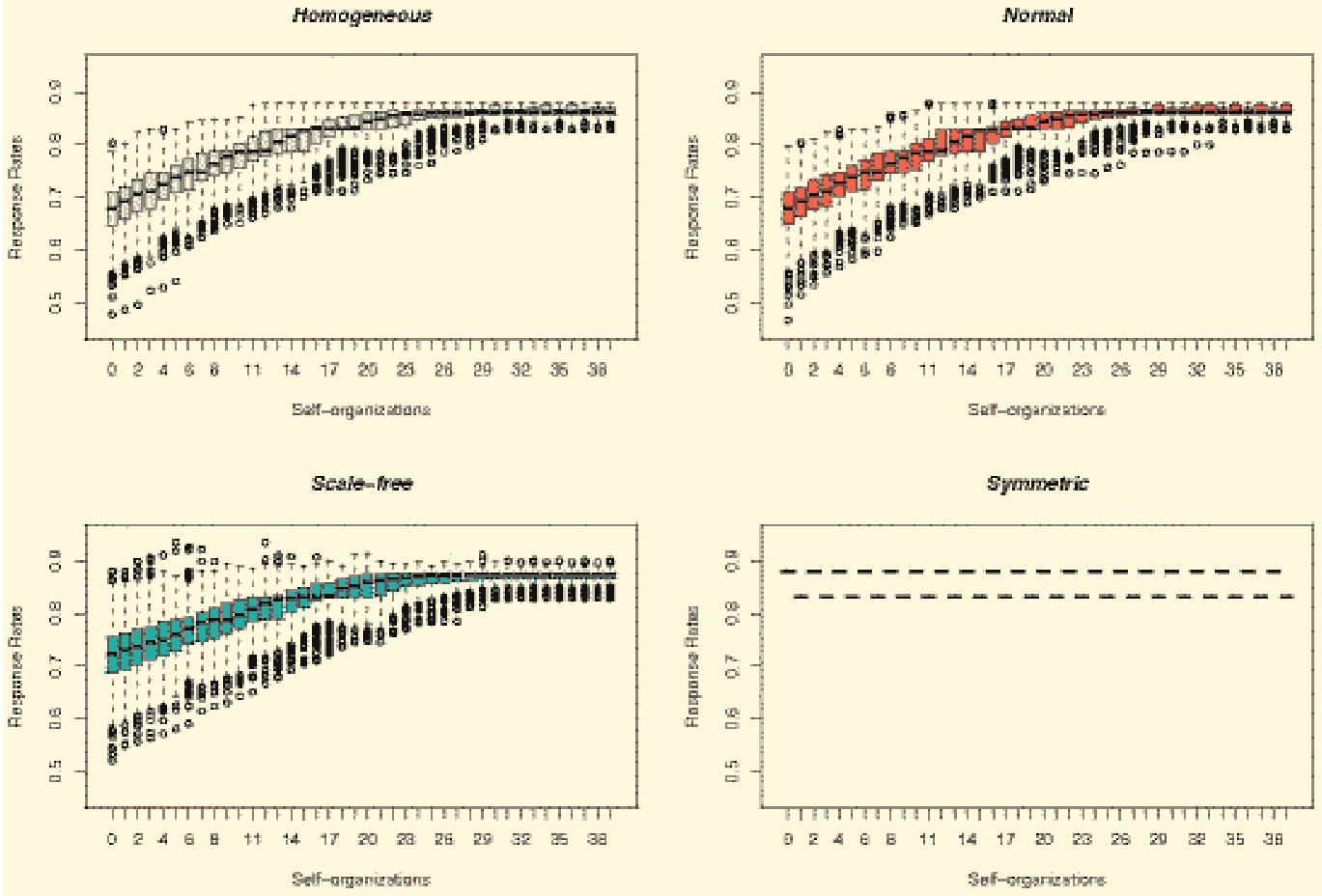} 
   \caption{Results for $N=15$, $K=15$.}
   \label{N15K15}
\end{figure}

As $K$ increases, it becomes more difficult to benefit from having several dependencies, since these will have also several dependencies. A high degree of connectivity also implies less hierarchy and less modularity. Thus, initially all topologies perform worse than symmetric, but through self-organization, they tend to reach a similar performance. This can be seen for the extreme case when $K=15$ (Figure \ref{N15K15}).

\begin{figure}[htbp] 
   \centering
   \includegraphics[width=6in]{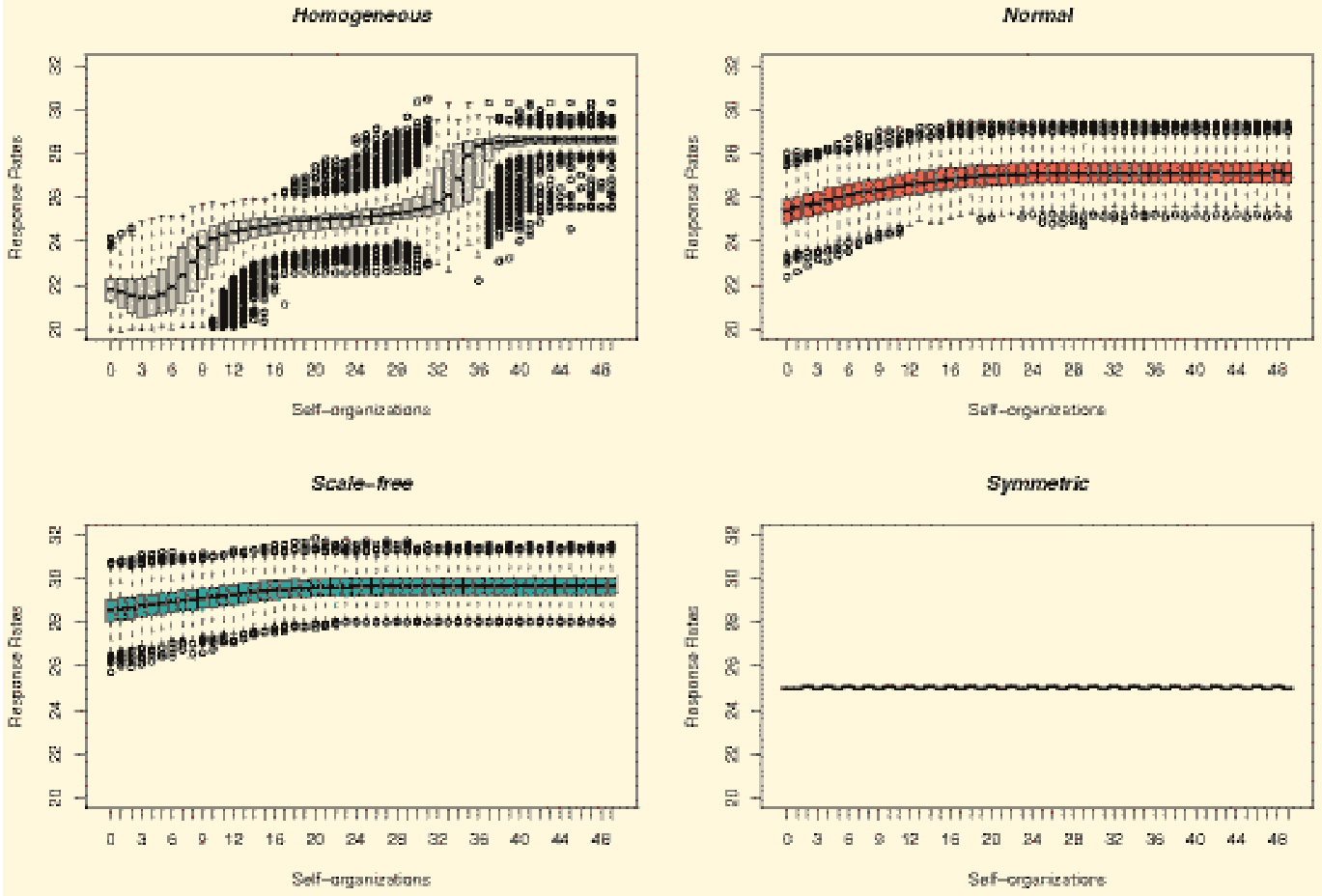} 
   \caption{Results for $N=100$, $K=1$.}
   \label{N100K1}
\end{figure}

\begin{figure}[htbp] 
   \centering
   \includegraphics[width=6in]{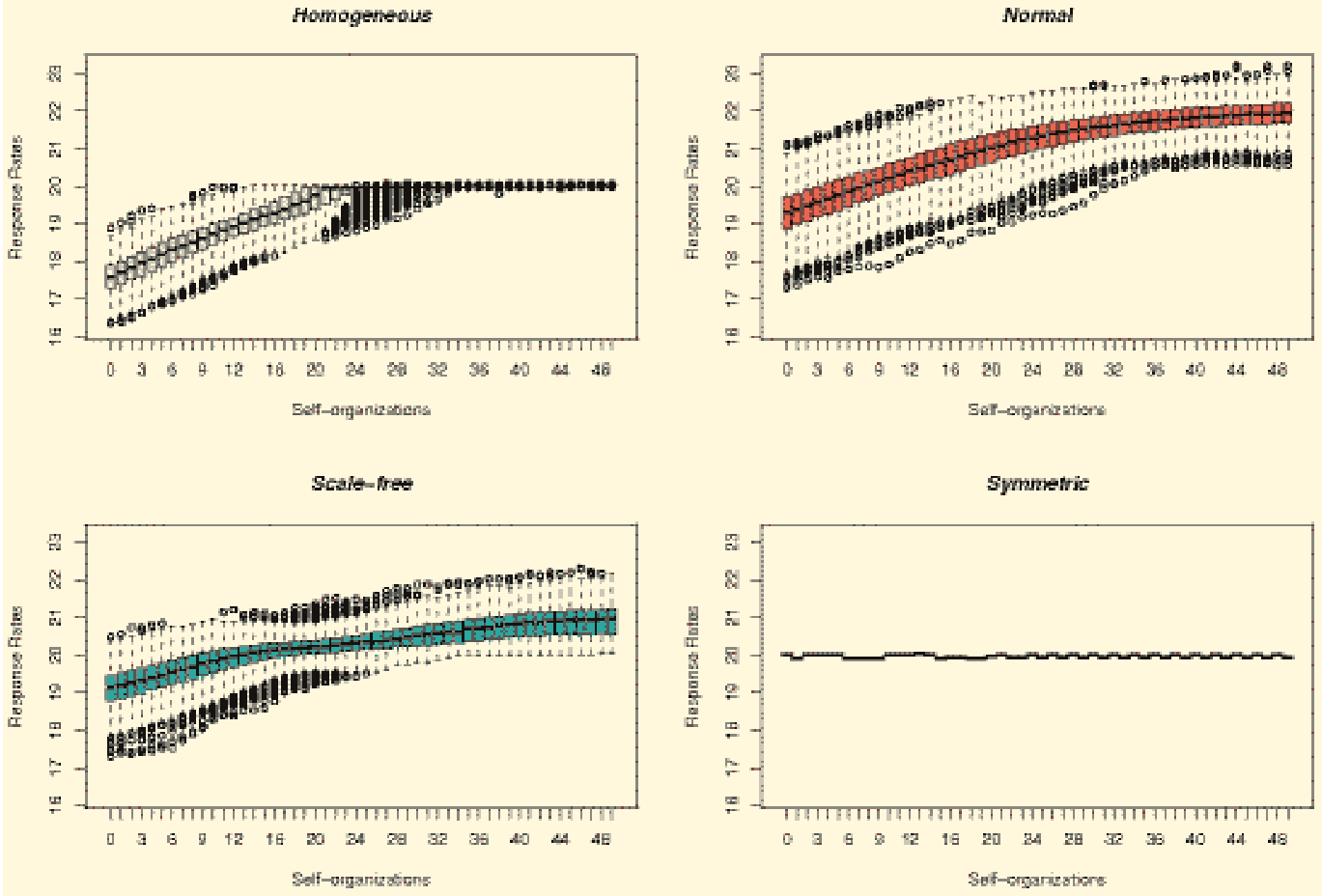} 
   \caption{Results for $N=100$, $K=2$.}
   \label{N100K2}
\end{figure}

\begin{figure}[htbp] 
   \centering
   \includegraphics[width=6in]{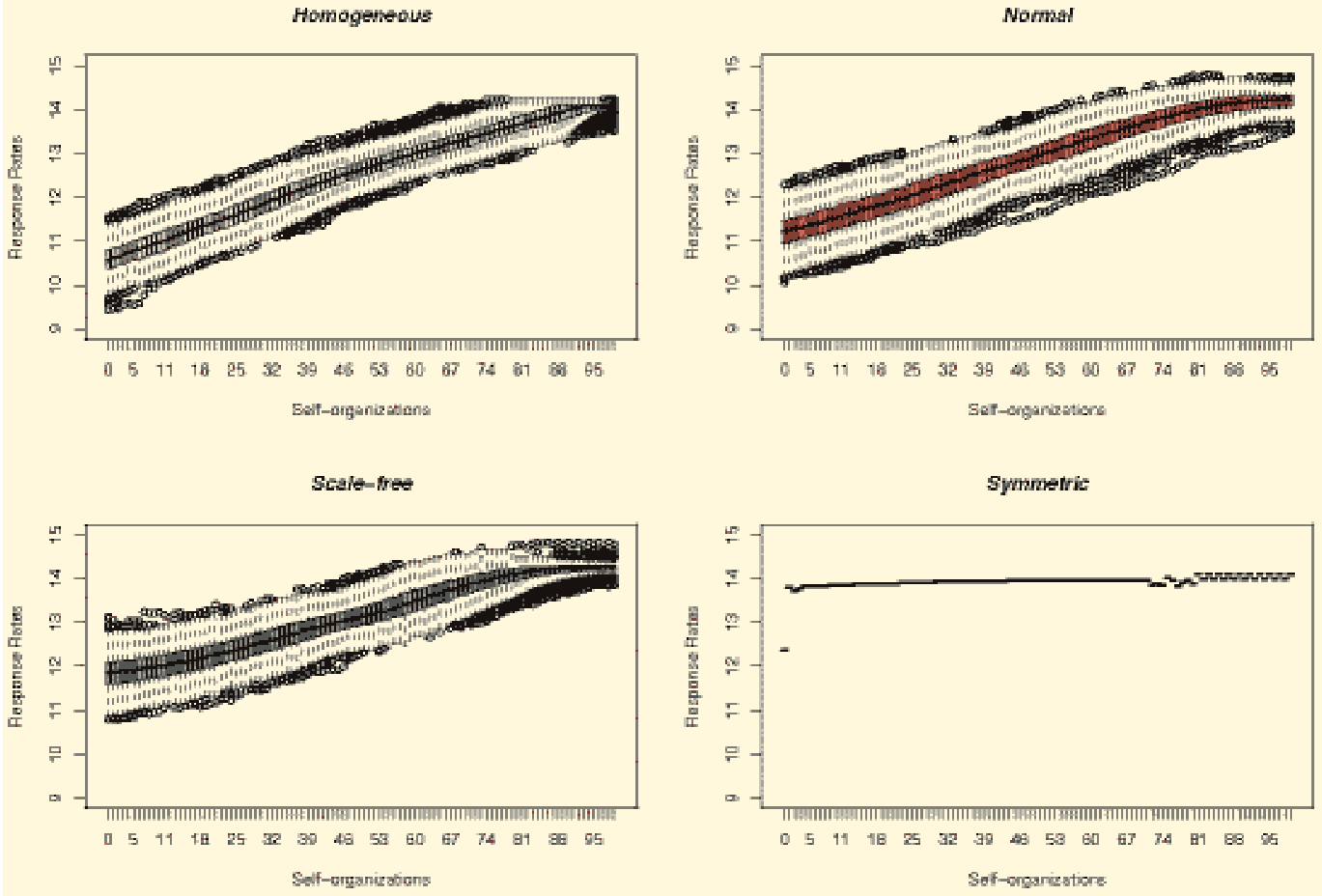} 
   \caption{Results for $N=100$, $K=5$.}
   \label{N100K5}
\end{figure}

\begin{figure}[htbp] 
   \centering
   \includegraphics[width=6in]{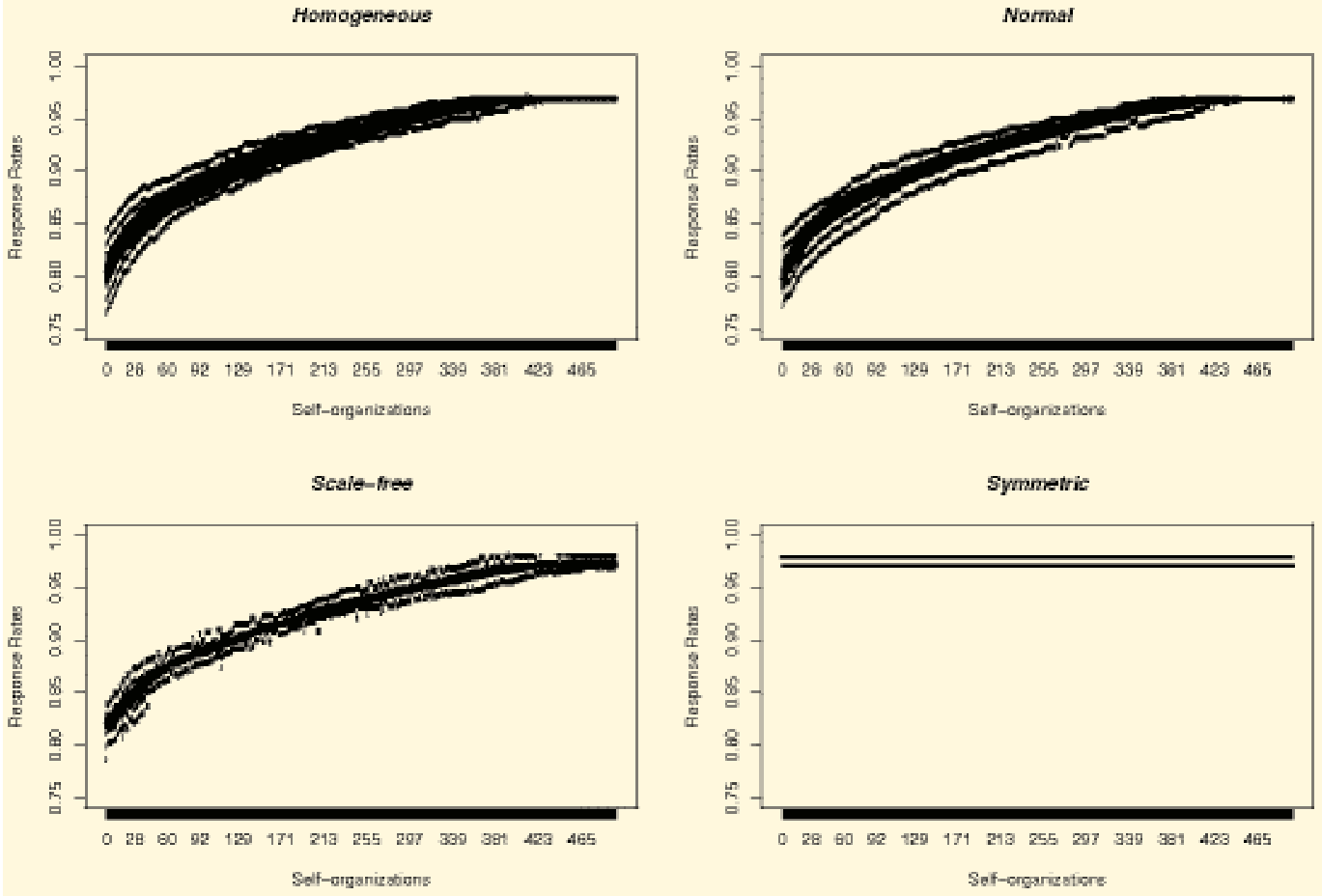} 
   \caption{Results for $N=100$, $K=100$.}
   \label{N100K100}
\end{figure}

The same behaviour as described above can  be seen for larger networks ($N=100$), for $K=1$ (Figure \ref{N100K1}), $K=2$ (Figure \ref{N100K2}), $K=5$ (Figure \ref{N100K5}), and $K=N$ (Figure \ref{N100K100}) (for this last case, only 25 networks were generated).

\subsection{RAN Discussion}

Many open questions remain in this model. However, the main goal was to illustrate the benefits of self-organization. The simulation results showed that only a few modifications of the network topology are required to increase performance to near optimal levels. For higher values of $K$, more self-organizations are required, simply because there are more dependencies in the network. Still, only a fraction of the total number of dependencies needs to be reconfigured to enable a \emph{random} network to achieve good performance.

Self-organization does not ensure optimality, but adaptability. For example, if the changes in demand of a bureaucracy change the decision time for a task, this can reconfigure itself to accommodate the change robustly. This can be useful for automatic detection of malfunctions and initial response to them: if an agent ``breaks down", its queue would grow, but the agents that have it as a dependency could rearrange their connectivity towards agents working properly.
Notice that the presented model assumes that all agents have equal decision and transmission delays. However, weights could be used to model diversity in the delays of agents.
Also, if in a real bureaucracy some dependencies cannot be changed, self-organization will find its way with the available flexible dependencies.

One remaining question is: when to stop self-organizing? In principle, self-organization can continue without degrading the performance of the network, while ensuring its adaptability. However, for the simple case presented here, there is no need of adaptation, so sooner or later the self-organization will take the RAN to a previously visited configuration. Some changes actually decrease slightly the performance, but the long run tendency is towards the highest possible performance for a particular network. If it is known what is the desired maximum performance, then that can be a criterium to stop the self-organization, but it is not obvious to know beforehand the maximum performance.

RAN-like models might also be useful to study organizational robustness, i.e. how well an organization can respond to node failure, using sensitivity analysis. For example, redundancy of nodes can be useful to ensure functionality of key or problematic tasks \cite{GershensonEtAl2006}.

\section{Conclusions}

This paper presented suggestions to use self-organizing techniques to
improve the efficiency of different aspects of bureaucracies. All the
improvements mentioned decrease different delays within a bureaucracy, reducing frictions and
leading to efficient adaptability and robustness. This is because increasing
speeds of reaction and decision will allow a bureaucracy to adapt quickly to
unexpected changes, while preserving its functionality. In consequence, the
``satisfactions" of agents, wether internal or external,
will be increased.

Standards and digital signatures certainly could be used to comply with the
formalities of bureaucracies. Using electronic media is not an impediment
for this. The adoption of these media is already underway, and it might bring in more benefits than just decreasing transmission delays. They can effectively support different types of self-organization within bureaucracies. The real value of self-organization will only be appreciated once it is applied in these organizations. But the ideas presented here are encouraging enough to try. 

Similar approaches could also be useful for other types of organizations: if frictions are reduced, satisfactions will increase.

\section{Acknowledgements}

I should like to thank Antonio Gershenson, Francis Heylighen, Mixel Kiemen, Diana Mangalagiu, and Marko Rodriguez
for useful discussions and relevant comments. This research was partly
supported by the Consejo Nacional de Ciencia y Teconolg\'{\i}a (CONACyT) of M\'{e}xico.

\bibliographystyle{apalike}
\bibliography{carlos,COG,orgs,sos,rbn}
\end{document}